# К теории рассеяния электронов на нейтронах

*Васильева А.А., Векленко Б.А.*

Процесс рассеяния электронов нейтронами определяет электропроводность и теплопроводность нейтронных звезд. Существующие теории дают заниженные значения этих величин.В настоящей работе мы обращаем внимание, что в случае резонансного или квазирезонансного рассеяние электрона на нейтроне стандартная теория возмущений не является достаточной. По мере приближения к резонансу рост сечения значительно превышает рост, предсказываемый теорией возмущений. В районе резонанса существует особенность теорией возмущений не описываемая. Излагаемый математический формализм заимствован из теории рассеяния света на возбужденных системах. Этот формализм принципиально использует свойства квантовых когерентных состояний электромагнитного поля. Ниже показано, каким образом можно воспользоваться этим методом расчета при описании рассеяния фермионов.

Ключевые слова: нейтроны, электроны, гамильтониан Ферми, теория возмущений, гигантский резонанс.

1. **Введение.**

В 1934 году для количественного описания спектра электронов в $\beta$ распаде Э. Ферми была предложена теория [1], построенная по аналогии с теорией спонтанной эмиссии возбужденного состояния атома [2]. Для выполнения закона сохранения энергии, Ферми ввел в теорию незаряженную безмассовую частицу –антинейтрино, существование которой было предсказано В.Паули. Теория Ферми получила дальнейшее развитие в виде теории слабого взаимодействия. Малая величина константы взаимодействия позволяет в теории $\beta$ распада ограничиться первым порядком теории возмущений. Рассеяние электрона на нейтроне требует в той же теории использования второго порядка теории возмущений.

В настоящей работе мы обращаем внимание, что в случае резонансного или квазирезонансного рассеяние электрона на нейтроне теория возмущений не является достаточной. По мере приближения к резонансу рост сечения значительно превышает рост, предсказываемый теорией возмущений. В районе резонанса существует особенность теорией возмущений не описываемая.

Настоящая работа посвящена выявлению этой особенности. Появление особенности не зависит от конкретной формы теории $\beta$ -распада. По этой причине в настоящем сообщении, в целях простоты, мы ограничимся исследованием нерелятивистского аналога теории Ферми и пренебрежем существованием антинейтрино.Исследуемый ниже гамильтониан можно рассматривать как модельный. Излагаемый математический формализм заимствован из теории рассеяния света на возбужденных системах [3]. Этот формализм принципиально использует свойства квантовых когерентных состояний[4] электромагнитного поля. Ниже показано, каким образом можно воспользоваться этим методом расчета при описании рассеяния фермионов.

Процесс рассеяния электронов нейтронами определяет электропроводность и теплопроводность нейтронных звезд . Существующие теории дают заниженные значения этих величин [5]. По этой причине выявление механизмов, ведущих к увеличению сечений рассеяния представляет актуальный интерес.

## 2. Математический формализм.

В методе вторичного квантования в представлении Шредингера сопоставим полю электронов полевой оператор $\hat{\psi}_e(\mathbf{r},t)$, заданный в любой момент времени $t$ в каждой точке пространства $\mathbf{r}$. Полю нейтронов сопоставим полевой оператор $\hat{\psi}_N(\mathbf{r},t)$, полю протонов - $\hat{\psi}_P(\mathbf{r},t)$. Спиновыми эффектами пренебрежем. Гамильтониан взаимодействия в представлении взаимодействия примем в виде

$$\hat{H}'(t) = q\int \hat{\psi}_P^+(\mathbf{r},t)\hat{\psi}_e^+(\mathbf{r},t)\hat{\psi}_N(\mathbf{r},t)d\mathbf{r} + H.c.$$

Этот гамильтониан описывает распад нейтрона на протон и электрон в пренебрежении существованием антинейтрино. Используемый гамильтониан можно рассматривать как модельный. Для наших целей он вполне достаточен, поскольку предсказываемый им аномальный резонанс в рассеянии оказывается типичным для многих систем. Вид полевых операторов совпадает с их выражениями в квантовой электродинамике. В частности

$$\hat{\psi}_e(x) = \frac{1}{\sqrt{V}}\sum_{\mathbf{p}} \hat{b}_{\mathbf{p}} \exp\left(i\frac{\mathbf{p}}{\hbar}\mathbf{r} - i\frac{\varepsilon(\mathbf{p})}{\hbar}t\right), \quad x = \{\mathbf{r},t\}. \, \varepsilon(\mathbf{p}) = \frac{p^2}{2m_e}. \qquad (1)$$

Здесь $\mathbf{p}$ -импульс электрона, $m_e$ -его масса, $V$ объем квантования. Через $\hat{b}_{\mathbf{p}}$ и $\hat{b}_{\mathbf{p}}^+$ обозначены операторы уничтожения и рождения электрона в состоянии с импульсом $\mathbf{p}$. Будем считать электроны подчиняющимися статистике Ферми –Дирака, и используем для них антикоммутационные соотношения

$$\hat{\psi}_e(\mathbf{r},t)\hat{\psi}_e^+(\mathbf{r}',t) + \hat{\psi}_e^+(\mathbf{r}',t)\hat{\psi}_e(\mathbf{r},t) = \left[\hat{\psi}_e(\mathbf{r},t);\hat{\psi}_e^+(\mathbf{r}',t)\right]_+ = \delta(\mathbf{r}-\mathbf{r}'),$$

вполне допустимыев нерелятивистской теории, несмотря на отсутствие спинов. Описание электронов в виде фермионных частиц для нас имеет принципиальное значение, поскольку ниже используется математический формализм, развитый ранее исключительно для описания бозонных полей. Использование этого формализма для фермионных полей, на первый взгляд, может казаться проблематичным. Будем считать нейтрон и протон, по каким либо причинам, локализованными в пространстве в районе координаты $\mathbf{R}=0$, и сопоставим им в представлении взаимодействия полевые операторы

$$\hat{\psi}_P(\mathbf{r},t) = \hat{b}_P \psi_P(\mathbf{r})\exp\left(-i\frac{\varepsilon_P}{\hbar}t\right), \quad \hat{\psi}_N(\mathbf{r},t) = \hat{b}_N \psi_N(\mathbf{r})\exp\left(-i\frac{\varepsilon_N}{\hbar}t\right), \qquad (2)$$

$$\int \psi_P^*(\mathbf{r})\psi_P(\mathbf{r})d\mathbf{r} = 1, \quad \int \psi_N^*(\mathbf{r})\psi_N(\mathbf{r})d\mathbf{r} = 1.$$

Через $\psi_P(\mathbf{r})$ и $\psi_N(\mathbf{r})$ обозначены волновые функции протона и нейтрона, $\varepsilon_P$ и $\varepsilon_N > \varepsilon_P$ ихэнергии, $\hat{b}_P$ и $\hat{b}_N$ -операторы уничтожения этих частиц,

$$\left[\hat{b}_P, \hat{b}_P^+\right]_+ = 1, \left[\hat{b}_N, \hat{b}_N^+\right]_+ = 1.$$

Что касается гамильтониана свободных полей $\hat{H}^0$, то примем его в стандартной форме в виде суммы гамильтонианов свободных полей

$$\hat{H}^0 = \hat{H}_e^0 + \hat{H}_P^0 + \hat{H}_N^0, \quad \hat{H}_e^0 = \sum_{\mathbf{p}} \varepsilon(\mathbf{p})\hat{b}_{\mathbf{p}}\hat{b}_{\mathbf{p}}^+, \quad \hat{H}_P^0 = \varepsilon_P \hat{b}_P^+ \hat{b}_P, \quad \hat{H}_N^0 = \varepsilon_N \hat{b}_N^+ \hat{b}_N, \quad .$$

Полный гамильтониан системы имеет вид

$$\hat{H} = \hat{H}^0 + \hat{H}'.$$

### 3. Стандартная теория возмущений.

Если в начальный момент времени в пространстве находился электрон в состоянии $\hat{b}_{\mathbf{p}_0}^+|0\rangle$ и один нейтрон, то волновая функция системы в представлении взаимодействия имела вид $|\Psi_0\rangle = \hat{b}_N^+ \hat{b}_{\mathbf{p}_0}^+|0\rangle$. Через $|0\rangle$ обозначена волновая функция вакуумного состояния системы. После процесса рассеяния по истечении времени $t \to \infty$ волновая функция системы $|\Psi(t)\rangle$ окажется равной

$$|\Psi(t)\rangle = \sum_{\mathbf{p}'} f_N(\mathbf{p}',t)\hat{b}_N^+ \hat{b}_{\mathbf{p}}^+|0\rangle + \sum_{\mathbf{p}_1 \mathbf{p}_2} f_P(\mathbf{p}_1,\mathbf{p}_2,t)\hat{b}_P^+ \hat{b}_{\mathbf{p}_1}^+ \hat{b}_{\mathbf{p}_2}^+|0\rangle, \quad f(\mathbf{p}_1,\mathbf{p}_2,t) = -f(\mathbf{p}_2,\mathbf{p}_1,t).$$

Согласно теории Дирака [6] $|f_N(\mathbf{p},t)|^2$ определяет условную вероятность $W_c(\mathbf{p})$ нахождения рассеянного электрона в состоянии с импульсом $\mathbf{p}$, при условии нахождения нейтрона в прежнем состоянии. Функция $|f_P(\mathbf{p}_1,\mathbf{p}_2,t)|^2 + |f_P(\mathbf{p}_2,\mathbf{p}_1,t)|^2$ определяет вероятность $W_n(\mathbf{p})$ нахождения после реакции одного электрона в состоянии с импульсом $\mathbf{p}_1$, другого в состоянии с импульсом $\mathbf{p}_2$. Нейтрон, при этом, переходит в протон. Принято называть первый канал рассеяния когерентным, второй –некогерентным. Безусловная вероятность нахождения одного электрона после рассеяния в состоянии с импульсом $\mathbf{p}$ определяется суммой

$|f_N(\mathbf{p},t)|^2 + \sum_{\mathbf{p}'}\left(|f_P(\mathbf{p},\mathbf{p}',t)|^2 + |f_P(\mathbf{p}',\mathbf{p},t)|^2\right)$. Нетрудно видеть, что усреднение оператора $\hat{b}_{\mathbf{p}}^+\hat{b}_{\mathbf{p}}$ по состоянию системы $|\Psi(t)\rangle$ имеет вид

$$\langle \Psi^+|\hat{b}_{\mathbf{p}}^+\hat{b}_{\mathbf{p}}|\Psi\rangle = |f_N(\mathbf{p},t)|^2 + 2\sum_{\mathbf{p}'}|f_P(\mathbf{p}_1,\mathbf{p}_2,t)|^2 = W_c(\mathbf{p}) + W_n(\mathbf{p}). \qquad (3)$$

Произведение операторов $\hat{b}_{\mathbf{p}}^+\hat{b}_{\mathbf{p}}$ носит название оператора числа электронов в состоянии с импульсом $\mathbf{p}$. Усредненное по состоянию $|\Psi(t)\rangle$ произведение $\hat{b}_{\mathbf{p}}^+\hat{b}_{\mathbf{p}}$ определяет " среднее " значение числа электронов в этом состоянии. Оператор полного числа частиц, очевидно, равен $\sum_{\mathbf{p}}\hat{b}_{\mathbf{p}}^+\hat{b}_{\mathbf{p}}$. Если же в системе присутствует лишь одна частица, то конструкция (3) в согласии с теорией Дирака определяет полную вероятность нахождения электрона в состоянии с импульсом $\mathbf{p}$. Эта конструкция достаточна для количественного описания процесса рассеяния.

Обратимся к теории возмущений. В представлении взаимодействия волновая функция системы $|\Psi(t)\rangle$ удовлетворяет уравнению [7]

$$i\hbar\frac{\partial|\Psi(t)\rangle}{\partial t} = \hat{H}'(t)|\Psi(t)\rangle.$$

Решение этого уравнения имеет вид

$$|\Psi(t)\rangle = \hat{S}|\Psi_0\rangle, \quad \hat{S} = \hat{T}\exp\left(\frac{1}{i\hbar}\int_{-\infty}^{t}\hat{H}(t')dt'\right),$$

Поскольку

$$\hat{S} = 1 + \hat{S}^{(1)} + \hat{S}^{(2)} + ..., \quad \hat{S}^{(n)} = \frac{\hat{T}}{n!}\left(\frac{1}{i\hbar}\int_{-\infty}^{\infty}\hat{H}'(t')dt'\right)^n,$$

то

$$\sum_{\mathbf{p}} f_N(\mathbf{p},t)\hat{b}_N^+\hat{b}_{\mathbf{p}}^+|0\rangle = \left(1 + \hat{S}^{(2)} + \hat{S}^{(4)} + ...\right)|\Psi_0\rangle, \sum_{\mathbf{p}_1\mathbf{p}_2} f_P(\mathbf{p}_1,\mathbf{p}_2,t)\hat{b}_N^+\hat{b}_{\mathbf{p}_1}^+\hat{b}_{\mathbf{p}_2}^+|0\rangle = \left(\hat{S}^{(1)} + \hat{S}^{(3)} + ...\right)|\Psi_0\rangle. \quad (4)$$

Такое представление является очевидным, так как оператор $\hat{H}'(t)$ содержит операторы $\hat{b}_N(t)$ и $\hat{b}_N^+(t)$ лишь в первой степени, и каждое воздействие оператора $\hat{\tilde{H}}(t)$ меняет нейтронное состояние системы на протонное и наоборот. Поскольку $\hat{S}^{(n)}(t) \propto q^n$, то низшее по заряду приближение вероятности процесса рассеяния $\propto q^4$ описывается оператором $\hat{S}^{(2)}(t)$. При этом, выражение $|f_N(\mathbf{p},t)|^2$ описывает вероятность рассеяния при условии, что в результате рассеяния система остается в нейтронном состоянии, протон в системе отсутствует. В этом приближении функцией $f_P(\mathbf{p}_1,\mathbf{p}_2,t)$ обычно пренебрегают. Но при изучении физических реакций нас интересуют не условные, а полные вероятности процессов рассеяния. Поэтому опускать некогерентный канал рассеяния при любой величине константы взаимодействия $q$, вообще говоря, нельзя. К тому же, при наличии в системе резонансных явлений теория возмущений вообще теряет смысл. Некоторые особенности резонансного рассеяния электрона на нейтроне удается описать аналитически.

Вычислим в явном виде операторы $\hat{S}^{(1)}(t)$ и $\hat{S}^{(2)}(t)$.

Начнем с оператора

$$\hat{S}^{(1)} = \frac{1}{i\hbar}\int_{-\infty}^{t}\hat{H}'(t')dt'. \quad (5)$$

При изучении стационарных процессов рассеяния допустимо [2] использовать предельный переход $t \to \infty$, что сильно упрощает расчеты. Мы воспользуемся этим приемом.

Подстановка (1) и (2) в (5) показывает, что

$$\hat{S}^{(1)} = -i\frac{2\pi q}{\hbar\sqrt{V}}\sum_{\mathbf{p}} d^*(\mathbf{p})\delta\left(\frac{\varepsilon_P + \varepsilon(\mathbf{p})}{\hbar} - \frac{\varepsilon_N}{\hbar}\right)\hat{b}_P^+\hat{b}_{\mathbf{p}}^+\hat{b}_N + H.c., \quad (6)$$

где

$$d(\mathbf{p}) = \int \psi_N^*(\mathbf{r}) \exp\left(i\frac{\mathbf{p}}{\hbar}\mathbf{r}\right)\psi_P(\mathbf{r})d\mathbf{r}$$

При вычислении оператора $\hat{S}^{(2)}(\infty)$ необходимо упростить конструкцию

$$\hat{S}^{(2)}(\infty) = \frac{\hat{T}}{2!}\left(\frac{1}{i\hbar}\int_{-\infty}^{\infty}\hat{H}'(t')dt'\right)^2 = \frac{\hat{T}q^2}{2!}\left(\int\hat{\psi}_P^+(x_1)\hat{\psi}_e^+(x_1)\hat{\psi}_N(x_1)dx_1 + \int\hat{\psi}_N^+(x_1)\hat{\psi}_e(x_1)\hat{\psi}_P(x_1)dx_1\right)\cdot$$
$$\cdot\left(\int\hat{\psi}_P^+(x_2)\hat{\psi}_e^+(x_2)\hat{\psi}_N(x_2)dx_2 + \int\hat{\psi}_N^+(x_2)\hat{\psi}_e(x_2)\hat{\psi}_P(x_2)dx_2\right), \quad dx = d\mathbf{r}dt. \tag{7}$$

Поскольку при вычислении (7) конструкции

$$\hat{\psi}_N(x_1)\hat{\psi}_N(x_2)|\Psi_0\rangle = \hat{\psi}_P(x_1)\hat{\psi}_P(x_2)|\Psi_0\rangle = 0 \tag{8}$$

обращаются в ноль, то в (7) нас интересуют лишь равные друг другу перекрестные члены

$$\hat{S}^{(2)} = \hat{T}q^2\int\hat{\psi}_N^+(x_1)\hat{\psi}_e(x_1)\hat{\psi}_P(x_1)dx_1\int\hat{\psi}_P^+(x_2)\hat{\psi}_e^+(x_2)\hat{\psi}_N(x_2)dx_2. \tag{9}$$

Переход от хронологического произведения операторов к их нормальному произведению, легко [8] осуществляется с помощью свертки полевых операторов

$$\left(\hat{T}-\hat{N}\right)\hat{\psi}_P(x_1)\hat{\psi}_P^+(x_2) = i\hbar G_P(x_1, x_2),$$

$$G_P(x_1, x_2) = \psi_P(x_1)\psi_P^+(x_2)\int_{-\infty}^{\infty}\frac{\exp(-iEt/\hbar)}{E-\varepsilon_P+i0}\frac{dE}{2\pi\hbar},$$

где $\hat{N}$ оператор нормального упорядочения полевых операторов. Учитывая равенства (8) и интересующую нас далее конструкцию $\hat{S}^{(2)}(\infty)|\Psi_0\rangle$, при вычислении (9) достаточно ограничиться одной сверткой протонных операторов

$$\hat{S}^{(2)} = i\hbar\hat{T}q^2\int\hat{\psi}_N^+(x_1)\hat{\psi}_e(x_1)G_P(x_1, x_2)\hat{\psi}_e^+(x_2)\hat{\psi}_N(x_2)dx_1 dx_2$$

После выполнения интегрирования по координатам и времени оказывается, что

$$\hat{S}^{(2)}(\infty) = -i\frac{2\pi q^2}{\hbar V}\sum_{\mathbf{p}'\mathbf{p}''}\frac{d(\mathbf{p}'')d^*(\mathbf{p}')\hat{b}_N^+\hat{b}_N\hat{b}_{\mathbf{p}'}^+\hat{b}_{\mathbf{p}''}}{\varepsilon_N-\varepsilon_P-\varepsilon_{\mathbf{p}''}+i0}\delta\left(\frac{\varepsilon_{\mathbf{p}'}-\varepsilon_{\mathbf{p}''}}{\hbar}\right). \tag{10}$$

Отсюда по известным правилам квантовой теории вероятность $w_c(\mathbf{p})$ когерентного рассеяния электрона в единицу времени в состояние с импульсом $\mathbf{p}$ оказывается равной

$$w_c(\mathbf{p}) = \frac{1}{2\pi}\left|\frac{2\pi q^2}{\hbar V}\frac{d(\mathbf{p}_0)d^*(\mathbf{p})}{\varepsilon_N-\varepsilon_P-\varepsilon_{\mathbf{p}_0}+i0} + H.c.\right|^2\delta\left(\frac{\varepsilon_\mathbf{p}-\varepsilon_{\mathbf{p}_0}}{\hbar}\right) \propto q^4. \tag{11}$$

Эта формула, очевидным образом, описывает стандартный резонанс в точке $\varepsilon(\mathbf{p}_0) = \varepsilon_N - \varepsilon_P$. Ниже показано, что полная вероятность $w_c(\mathbf{p}) + w_n(\mathbf{p})$ в этой точке с учетом некогерентного канала рассеяния $w_n(\mathbf{p})$ обладает значительно более ярко выраженной особенностью.

Эту особенность, казалось бы, можно изучить, рассматривая возникающую в некогерентном канале часть конструкции

$$\left\langle \Psi_0^+ \left| \left( \hat{S}^{(1)} + \hat{S}^{(3)} \right)^+ \hat{b}_{\mathbf{p}}^+ \hat{b}_{\mathbf{p}} \left( \hat{S}^{(1)} + \hat{S}^{(3)} \right) \right| \Psi_0 \right\rangle, \qquad (12)$$

пропорциональную $q^4$

$$\left\langle \Psi_0^+ \left| \hat{S}^{(1)+} \hat{b}_{\mathbf{p}}^+ \hat{b}_{\mathbf{p}} \hat{S}^{(3)} \right| \Psi_0 \right\rangle + \left\langle \Psi_0^+ \left| \hat{S}^{(3)+} \hat{b}_{\mathbf{p}}^+ \hat{b}_{\mathbf{p}} \hat{S}^{(1)} \right| \Psi_0 \right\rangle.$$

Но такая сумма не является положительно определенной, и нарушает положительную определенность некогерентного канала в целом. Рассмотрение такой суммы по этой причине теряет смысл. Положительная определенность восстанавливается путем добавления слагаемых, содержащих операторы $\hat{S}^{(1)}$ и $\hat{S}^{(3)+}$, что возвращает нас к (12). Таким образом, содержащиеся в (12) слагаемые, пропорциональные $q^6$, играют определяющую роль. Далее заметим, что учтенные указанным образом члены пропорциональные $q^6$, не исчерпывают всех слагаемых $\propto q^6$ в некогерентном канале. Такие же степени заряда возникают в высшем приближении, определяемом произведением $\hat{S}^{(1)} \hat{S}^{(5)+}$. И так далее. Таким образом, стандартным образом построить теорию возмущений по заряду на основе формулы (3) не представляется возможным. Суммы бесконечных подпоследовательностей членов с более высокими степенями могут описываться выражениями, обладающими более низкими степенями заряда. Таковы известные свойства бесконечных рядов в теории квантованных полей. К этому вопросу ниже мы вернемся еще раз. Возникает вопрос: нельзя ли перестроить теорию так, чтобы вклад слагаемых четвертой степени в конструкцию $\left\langle \Psi_0^+ \left| \hat{S}^+ \hat{b}_{\mathbf{p}}^+ \hat{b}_{\mathbf{p}} \hat{S} \right| \Psi^0 \right\rangle$ был бы положительно определен.

### 4. Метод когерентных амплитуд

Ниже излагается метод, позволяющий уточнить формулу (11). Этот метод, уточняющий стандартную теорию возмущений, заимствован из теории рассеяния частиц Бозе–Эйнштейна [3]. Ниже предлагается модификация этого метода применительно к частицам Ферми–Дирака.

Если в начальном состоянии исследуемая система обладала одним нейтроном и одним рассеиваемом на нем электроном, находящимся в состоянии с импульсом $\mathbf{p}_0$, то волновая функция системы имела вид $\left| \Psi_0 \right\rangle = \hat{b}_N^+ \hat{b}_{\mathbf{p}_0}^+ \left| 0 \right\rangle$. Если же в начальном состоянии электрон отсутствовал, то $\left| \Psi_0 \right\rangle = \hat{b}_N^+ \left| 0 \right\rangle$. Суперпозиция волновых функций $\left| \Psi_0 \right\rangle \propto \left( 1 + b_{\mathbf{p}_0} \hat{b}_{\mathbf{p}_0}^+ \right) \hat{b}_N^+ \left| 0 \right\rangle$ вновь является волновой функцией системы, но с неопределенным числом электронов. Такой вид начального состояния системы является возможным, и для дальнейшего будет иметь принципиальное значение. Предварительно пронормируем волновую функцию $\Psi_0$ на единицу. Так как

$$\left\langle \Psi_0^+ \Psi_0 \right\rangle = \left\langle 0 \left| \hat{b}_N \left( 1 + b_{\mathbf{p}_0}^* \hat{b}_{\mathbf{p}_0} \right) \left( 1 + b_{\mathbf{p}_0} \hat{b}_{\mathbf{p}_0}^+ \right) \hat{b}_N^+ \right| 0 \right\rangle = 1 + \left| b_{\mathbf{p}_0} \right|^2,$$

То искомая волновая функция имеет вид

$$|\Psi_0\rangle = \frac{\left(1+b_{\mathbf{p}_0}\hat{b}^+_{\mathbf{p}_0}\right)\hat{b}^+_N|0\rangle}{\sqrt{1+\left|b_{\mathbf{p}_0}\right|^2}}. \tag{13}$$

" Среднее " число частиц в системе при этом равно

$$\langle\Psi_0^+|\sum_{\mathbf{p}}\hat{b}^+_{\mathbf{p}}\hat{b}_{\mathbf{p}}|\Psi_0\rangle = \langle 0|\frac{\left(1+\hat{b}^*_{\mathbf{p}_0}\hat{b}_{\mathbf{p}}\right)\hat{b}_N}{\sqrt{1+\left|b_{\mathbf{p}_0}\right|^2}}\sum_{\mathbf{p}}\hat{b}^+_{\mathbf{p}}\hat{b}_{\mathbf{p}}\frac{\left(1+b_{\mathbf{p}_0}\hat{b}^+_{\mathbf{p}_0}\right)\hat{b}^+_N}{\sqrt{1+\left|b_{\mathbf{p}_0}\right|^2}}|0\rangle = \frac{\left|b_{\mathbf{p}_0}\right|^2}{1+\left|b_{\mathbf{p}_0}\right|^2}.$$

Волновая функция (13) построена по аналогии с волновой функцией когерентного состояния фотонного поля [4], и обладает похожими математическими свойствами. В частности, в этом состоянии квантовое среднее оператора $\hat{\psi}_e(\mathbf{r},t)$ отлично от нуля

$$\langle\Psi_0^+|\hat{\psi}_e(\mathbf{r},t)|\Psi_0\rangle = \frac{b_{\mathbf{p}_0}}{1+\left|b_{\mathbf{p}_0}\right|^2}\frac{1}{\sqrt{V}}\exp\left(-i\frac{\varepsilon(\mathbf{p})}{\hbar}t+i\frac{\mathbf{p}_0}{\hbar}\mathbf{r}\right),$$

Нас будет интересовать конструкция $\langle\Psi^+|\hat{b}^+_{\mathbf{p}}\hat{b}_{\mathbf{p}}|\Psi\rangle$, где усреднение проводится по состоянию системы после процесса рассеяния. " Среднее " число частиц в состоянии с импульсом $\mathbf{p}$, если в начальный момент времени частица обладала импульсом $\mathbf{p}_0$, определяет собой вследствие рассеяния вероятность $W(\mathbf{p})$ перехода электрона из начального состояния в конечное.

Аналогично представлению (4), согласно определению имеем

$$\langle\Psi^+|\hat{b}^+_{\mathbf{p}}\hat{b}_{\mathbf{p}}|\Psi\rangle = \langle\hat{b}^+_{\mathbf{p}}\hat{b}_{\mathbf{p}}\rangle = \langle\Psi_0^+|\hat{S}^+\hat{b}^+_{\mathbf{p}}\hat{b}_{\mathbf{p}}\hat{S}|\Psi_0\rangle = \langle\Psi_0^+|\hat{S}^+\hat{b}^+_{\mathbf{p}}\hat{b}_{\mathbf{p}}\hat{S}|\Psi_0\rangle_c + \langle\Psi_0^+|\hat{S}^+\hat{b}^+_{\mathbf{p}}\hat{b}_{\mathbf{p}}\hat{S}|\Psi_0\rangle_n,$$

где

$$\langle\Psi_0^+|\hat{S}^+\hat{b}^+_{\mathbf{p}}\hat{b}_{\mathbf{p}}\hat{S}|\Psi_0\rangle_c = \langle\hat{b}^+_{\mathbf{p}}\hat{b}_{\mathbf{p}}\rangle_c = \langle\Psi_0^+|\left(1+\hat{S}^{(2)}+...\right)^+\hat{b}^+_{\mathbf{p}}\hat{b}_{\mathbf{p}}\left(1+\hat{S}^{(2)}+...\right)|\Psi_0\rangle,$$

$$\langle\Psi_0^+|\hat{S}^+\hat{b}^+_{\mathbf{p}}\hat{b}_{\mathbf{p}}\hat{S}|\Psi_0\rangle_n = \langle\hat{b}^+_{\mathbf{p}}\hat{b}_{\mathbf{p}}\rangle_c = \langle\Psi_0^+|\left(\hat{S}^{(1)}+\hat{S}^{(3)}...\right)^+\hat{b}^+_{\mathbf{p}}\hat{b}_{\mathbf{p}}\left(\hat{S}^{(1)}+\hat{S}^{(3)}...\right)|\Psi_0\rangle,$$

$$\langle\hat{b}^+_{\mathbf{p}}\hat{b}_{\mathbf{p}}\rangle = \langle\hat{b}^+_{\mathbf{p}}\hat{b}_{\mathbf{p}}\rangle_c + \langle\hat{b}^+_{\mathbf{p}}\hat{b}_{\mathbf{p}}\rangle_n.$$

Будем говорить о когерентном канале рассеяния, при котором в результате рассеяния нейтрон остается в прежнем состоянии. В результате некогерентного рассеяния нейтрон заменяется протоном.

Пусть согласно определению

$$\langle\hat{b}_{\mathbf{p}}\rangle_c = \langle\Psi_0^+|\left(1+\hat{S}^{(2)}+...\right)^*\hat{b}_{\mathbf{p}}\left(1+\hat{S}^{(2)}+...\right)|\Psi_0\rangle,\ \langle\hat{b}^+_{\mathbf{p}}\rangle_c = \langle\hat{b}_{\mathbf{p}}\rangle^*_c, \tag{14}$$

$$\langle\hat{b}_{\mathbf{p}}\rangle_n = \langle\Psi_0^+|\left(\hat{S}^{(1)}+\hat{S}^{(3)}+...\right)^*\hat{b}_{\mathbf{p}}\left(\hat{S}^{(1)}+\hat{S}^{(3)}+...\right)|\Psi_0\rangle,\ \langle\hat{b}^+_{\mathbf{p}}\rangle_n = \langle\hat{b}_{\mathbf{p}}\rangle^*_n. \tag{15}$$

Воспользовавшись очевидным неравенством

$$\left\langle \left(\hat{b}_{\mathbf{p}}^+ - \left\langle \hat{b}_{\mathbf{p}}^+ \right\rangle\right)\left(\hat{b}_{\mathbf{p}} - \left\langle \hat{b}_{\mathbf{p}} \right\rangle\right)\right\rangle_c \geq 0,$$

находим, что

$$\left\langle \hat{b}_{\mathbf{p}}^+ \hat{b}_{\mathbf{p}} \right\rangle_c \geq \left\langle \hat{b}_{\mathbf{p}}^+ \right\rangle_c \left\langle \hat{b}_{\mathbf{p}} \right\rangle_c. \tag{16}$$

Аналогично

$$\left\langle \hat{b}_{\mathbf{p}}^+ \hat{b}_{\mathbf{p}} \right\rangle_n \geq \left\langle \hat{b}_{\mathbf{p}}^+ \right\rangle_n \left\langle \hat{b}_{\mathbf{p}} \right\rangle_n. \tag{17}$$

Таким образом

$$\left\langle \hat{b}_{\mathbf{p}}^+ \hat{b}_{\mathbf{p}} \right\rangle \geq \left\langle \hat{b}_{\mathbf{p}}^+ \right\rangle_c \left\langle \hat{b}_{\mathbf{p}} \right\rangle_c + \left\langle \hat{b}_{\mathbf{p}}^+ \right\rangle_n \left\langle \hat{b}_{\mathbf{p}} \right\rangle_n.$$

Это, не связанное с теорией возмущений, неравенство дает возможность оценить снизу сечение рассеяния. В низшем порядке теории возмущений из (14) и (10) следует

$$\left\langle \hat{b}_{\mathbf{p}} \right\rangle_c = \left\langle \Psi_0^* \middle| \hat{b}_{\mathbf{p}} \hat{S}^{(2)} \middle| \Psi_0 \right\rangle = \left\langle \hat{b}_{\mathbf{p}} \right\rangle_c = -i \frac{2\pi q^2}{\hbar^2 V} \frac{d(\mathbf{p}_0) d^*(\mathbf{p})}{\varepsilon_N - \varepsilon_P - \varepsilon_{\mathbf{p}_0} + i0} \frac{b_{\mathbf{p}_0}}{1 + \left|b_{\mathbf{p}_0}\right|^2} \delta\left(\frac{\varepsilon_{\mathbf{p}} - \varepsilon_{\mathbf{p}_0}}{\hbar}\right) \tag{18}$$

Аналогично, из (15) посредством (6) имеем

$$\left\langle \hat{b}_{\mathbf{p}} \right\rangle_n = \left\langle \hat{S}^{(1)+} \hat{b}_{\mathbf{p}} \hat{S}^{(1)} \right\rangle_0 = \frac{q^2}{\hbar^2} \frac{(2\pi)^2}{V} \delta\left(\frac{\varepsilon_P + \varepsilon_{\mathbf{p}_0} - \varepsilon_N}{\hbar}\right) d(\mathbf{p}_0) d^*(\mathbf{p}) \frac{b_{\mathbf{p}_0}}{1 + \left|b_{\mathbf{p}_0}\right|^2} \delta\left(\frac{\varepsilon(\mathbf{p}) - \varepsilon(\mathbf{p}_0)}{\hbar}\right). \tag{19}$$

## 5. Сечение рассеяния

«Среднее» число частиц в начальном состоянии, описываемом волновой функцией (13), равно

$$\left\langle \Psi_0 \middle| \hat{b}_{\mathbf{p}_0}^+ \hat{b}_{\mathbf{p}_0} \middle| \Psi_0 \right\rangle = \frac{\left|b_{\mathbf{p}_0}\right|^2}{1 + \left|b_{\mathbf{p}_0}\right|^2}.$$

Среднее число частиц в состоянии с импульсом $\mathbf{p} \neq \mathbf{p}_0$ после рассеяния определяется конструкцией $\left\langle \hat{b}_{\mathbf{p}}^+ \hat{b}_{\mathbf{p}} \right\rangle$. Вероятность перехода $W(\mathbf{p})$ одной частицы из начального состояния в конечное при $t \to \infty$, таким образом, равна

$$W(\mathbf{p}) = \left\langle \hat{b}_{\mathbf{p}}^+ \hat{b}_{\mathbf{p}} \right\rangle \frac{1 + \left|b_{\mathbf{p}_0}\right|^2}{\left|b_{\mathbf{p}_0}\right|^2}.$$

Формулы (16-19) показывают, что

$$W_c(\mathbf{p}) \geq \left| \frac{2\pi q^2}{\hbar V} \frac{d(\mathbf{p}_0,\mathbf{k}) d^*(\mathbf{p},\mathbf{k})}{\varepsilon_N - \varepsilon_P - \varepsilon(p_0) + i0} \delta\left( \frac{\varepsilon_{\mathbf{p}} - \varepsilon_{\mathbf{p}_0}}{\hbar} \right) \right|^2,$$

$$W_n(\mathbf{p}) \geq \left| \frac{q^2}{\hbar^2} \frac{(2\pi)^2}{V} \delta\left( \frac{\varepsilon_N - \varepsilon_P - \varepsilon(p_0)}{\hbar} \right) d(\mathbf{p}_0,\mathbf{k}) d^*(\mathbf{p},\mathbf{k}) \delta\left( \frac{\varepsilon_{\mathbf{p}} - \varepsilon_{\mathbf{p}_0}}{\hbar} \right) \right|^2. \quad (20)$$

Согласно правилам квантовой теории полная вероятность перехода в единицу времени $w(\mathbf{p})$ в низшем порядке теории возмущений равна

$$w(\mathbf{p}) = w_c(\mathbf{p}) + w_n(\mathbf{p}) \geq \frac{1}{2\pi} \left| \frac{2\pi q^2}{\hbar V} \frac{d(\mathbf{p}_0) d^*(\mathbf{p})}{\varepsilon_N - \varepsilon_P - \varepsilon(p_0) + i0} \right|^2 \delta\left( \frac{\varepsilon_{\mathbf{p}} - \varepsilon_{\mathbf{p}_0}}{\hbar} \right) +$$

$$+ \frac{1}{2\pi} \left| \frac{q^2}{\hbar^2} \frac{(2\pi)^2}{V} \delta\left( \frac{\varepsilon_N - \varepsilon_P - \varepsilon(p_0)}{\hbar} \right) d(\mathbf{p}_0) d^*(\mathbf{p}) \right|^2 \delta\left( \frac{\varepsilon_{\mathbf{p}} - \varepsilon_{\mathbf{p}_0}}{\hbar} \right). \quad (21)$$

Первое слагаемое этой формулы повторяет результат (11) стандартной теории рассеяния. Второе слагаемое указывает на то, что в резонансной области $\varepsilon(p_0) = \varepsilon_N - \varepsilon_P$ согласно некогерентному каналу рассеяния дополнительно появляется сингулярное слагаемое, существенно увеличивающее сечение процесса рассеяния.

Такой рост вероятности рассеяния наглядно проявляет себя при учете энергетических ширин энергетических уровней рассеивателя, учитываемый параметром $\gamma$. Такой учет деформирует формулу (21) следующим образом

$$w(\mathbf{p}) = w_c(\mathbf{p}) + w_n(\mathbf{p}) \geq \frac{1}{2\pi} \left| \frac{2\pi q^2}{\hbar V} \frac{d(\mathbf{p}_0) d^*(\mathbf{p})}{\varepsilon_N - \varepsilon_P - \varepsilon(p_0) + i\gamma/2} \right|^2 \delta\left( \frac{\varepsilon_{\mathbf{p}} - \varepsilon_{\mathbf{p}_0}}{\hbar} \right) +$$

$$+ \frac{1}{2\pi} \left| \frac{q^2}{\hbar} \frac{2\pi}{V} \frac{\gamma}{(\varepsilon_N - \varepsilon_P - \varepsilon(p_0))^2 + \gamma^2/4} d(\mathbf{p}_0) d^*(\mathbf{p}) \right|^2 \delta\left( \frac{\varepsilon_{\mathbf{p}} - \varepsilon_{\mathbf{p}_0}}{\hbar} \right).$$

Теперь ясно видно, что в случае точного резонанса $\varepsilon(p_0) = \varepsilon_N - \varepsilon_P$ вероятность рассеяния в некогерентном канале вчетверо превосходит вероятность когерентного рассеяния электрона

$$w_n(\mathbf{p}) = 4 w_n(\mathbf{p}).$$

Таким образом, учет некогерентного канала рассеяния полное сечение рассеяния увеличивает в пять раз

$$w(\mathbf{p}) = w_c(\mathbf{p}) + w_n(\mathbf{p}) = 5 w_c(\mathbf{p}). \quad (22)$$

Вероятность рассеяния электрона в единицу времени $p(\mathbf{p})$ определяет дифференциальное сечение рассеяния $\sigma(\mathbf{p})$ согласно следующей формуле

$$\sigma(\mathbf{p}) = \frac{m_e}{p_0} \int_0^\infty w(\mathbf{p}) \frac{V^2}{(2\pi\hbar)^3} p^2 dp.$$

Поскольку $w(\mathbf{p}) \propto V^{-2}$, то нормировочный объем $V$ из результата выпадает. Согласно (22) учет некогерентного канала рассеяния увеличивает в условиях резонанса сечение рассеяния в пять раз.

### 6. Перестройка ряда теории возмущений

Как следует из предыдущего параграфа изменение структуры начального квантового состояния электрона позволяет предвидеть дополнительную особенность в резонансном рассеянии электрона на нейтроне. Возникает естественный вопрос о том, как получить аналогичный результат при начальном состоянии электрона, заданным в стандартной фоковской форме $\left|\Psi_0\right\rangle = \hat{b}_N^+ \hat{b}_{\mathbf{p}_0}^+ \left|0\right\rangle$. Ниже решается эта задача путем перестройки ряда теории возмущений. Будет получен результат предыдущего раздела. Тем не менее, предлагаемый ниже вывод, формально правильный, не представляется убедительным. В математически корректной теории результат расчета не должен зависеть от формы ряда теории возмущений. В квантовой теории такое оказывается возможным, поскольку при любой константе взаимодействия ряд теории возмущений здесь расходящийся. По этой причине важно отметить, что вывод, предложенный в предыдущем параграфе, подтверждает правильность полученного ниже результата.

Нас интересует некогерентный канал рассеяния. Воспользовавшись тождествами

$$\sum_i \left|\Psi_0^{(i)}\right\rangle\left\langle\Psi_0^{(i)+}\right| = 1, \quad \sum_i \left|\hat{S}\Psi_0^{(i)}\right\rangle\left\langle\Psi_0^{(i)+}\hat{S}^+\right| = 1,$$

$$\hat{b}_{\mathbf{p}}^+ \sum_i \left|\Psi_0^{(i)}\right\rangle\left\langle\Psi_0^{(i)+}\right| \hat{b}_{\mathbf{p}} = \hat{b}_{\mathbf{p}}^+ \hat{b}_{\mathbf{p}}, \quad \hat{b}_{\mathbf{p}}^+ \sum_i \left|\hat{S}\Psi_0^{(i)}\right\rangle\left\langle\Psi_0^{(i)+}\hat{S}^+\right| \hat{b}_{\mathbf{p}} = \hat{b}_{\mathbf{p}}^+ \hat{b}_{\mathbf{p}},$$

где $\left|\Psi_0^{(i)}\right\rangle$-собственные функции свободного гамильтониана $\hat{H}^0$, найдем, что при $\mathbf{p} \neq \mathbf{p}_0$

$$\left\langle \hat{b}_{\mathbf{p}}^+ \hat{b}_{\mathbf{p}} \right\rangle_n = \left\langle \Psi_0^+ \left| \left(\hat{S}^{(1)} + \hat{S}^{(3)} + ...\right)^+ \hat{b}_{\mathbf{p}}^+ \hat{b}_{\mathbf{p}} \left(\hat{S}^{(1)} + \hat{S}^{(3)} + ...\right) \right| \Psi_0 \right\rangle =$$

$$= \sum_i \left| \left\langle \Psi_0^{(i)+} \left| \hat{S}^+ \hat{b}_{\mathbf{p}} \left(\hat{S}^{(1)} + \hat{S}^{(3)} + ...\right) \right| \Psi_0 \right\rangle \right|^2, \qquad (23)$$

$$\left\langle \Psi_0^+ \left| \hat{S}^{(1)+} \hat{b}_{\mathbf{p}}^+ \hat{b}_{\mathbf{p}} \hat{S}^{(1)} \right| \Psi_0 \right\rangle = \left\langle \Psi_0^+ \left| \hat{S}^{(1)+} \hat{b}_{\mathbf{p}}^+ \sum_i \left|\Psi_0^{(i)}\right\rangle\left\langle\Psi_0^{(i)+}\right| \hat{b}_{\mathbf{p}} \hat{S}^{(1)} \right| \Psi_0 \right\rangle = \sum_i \left| \left\langle \Psi_0^{(i)+} \left| \hat{b}_{\mathbf{p}} \hat{S}^{(1)} \right| \Psi_0 \right\rangle \right|^2.$$

Последняя строчка описывает вероятность спонтанного распада нейтрона, рассчитанная в работе [1]. Исключая спонтанное излучение, для вероятности рассеяния электрона в низшем порядке теории возмущений имеем

$$W_n(\mathbf{p}) = \left\langle \hat{b}_{\mathbf{p}}^+ \hat{b}_{\mathbf{p}} \right\rangle_n - \left\langle \Psi_0^+ \left| \hat{S}^{(1)+} \hat{b}_{\mathbf{p}}^+ \hat{b}_{\mathbf{p}} \hat{S}^{(1)} \right| \Psi_0 \right\rangle = \sum_i \left| \left\langle \Psi_0^{(i)+} \left| \hat{S}^{(1)+} \hat{b}_{\mathbf{p}} \hat{S}^{(1)} \right| \Psi_0 \right\rangle \right|^2 \propto q^4. \qquad (24)$$

Мы воспользовались тем, что

$$\left\langle \Psi_0^{(i)+} \left| \hat{S}^{(1)+} \hat{b}_\mathbf{p} \hat{S}^{(1)} \right| \Psi_0 \right\rangle \left\langle \Psi_0^{(i)+} \left| \hat{b}_\mathbf{p} \hat{S}^{(1)} \right| \Psi_0 \right\rangle^* = 0$$

при любой $\left|\Psi_0^{(i)}\right\rangle$. В формуле (24) опущено пропорциональное $q^4$ слагаемое.

$$\left\langle \Psi_0^{(i)+} \left| \hat{S}^{(2)+} \hat{b}_\mathbf{p} \hat{S}^{(1)} \right| \Psi_0 \right\rangle \left\langle \Psi_0^{(i)+} \left| \hat{b}_\mathbf{p} \hat{S}^{(1)} \right| \Psi_0 \right\rangle^*$$

Это слагаемое описывает поправку к полю электронов $\left|\Psi_0^{(i)}\right\rangle$, формируемому спонтанным распадом нейтрона, и к процессам рассеяния электронов отношения не имеет.

Из формулы (23), для вероятности $P_n(\mathbf{p})$ с учетом (6) находим

$$W_n(\mathbf{p}) \geq \left| \frac{q^2}{\hbar^2} \frac{(2\pi)^2}{V} \delta\left( \frac{\varepsilon_N - \varepsilon_P - \varepsilon(p_0)}{\hbar} \right) d(\mathbf{p}_0, \mathbf{k}) d^*(\mathbf{p}, \mathbf{k}) \delta\left( \frac{\varepsilon_\mathbf{p} - \varepsilon_{\mathbf{p}_0}}{\hbar} \right) \right|^2$$

Структура этой формулы повторяет структуру формулы (20), подтверждая тем самым ее справедливость.

### 7. Заключение

В настоящей работе рассеяние электрона на нейтроне описывается посредством перестроенного ряда теории возмущений. Перестроенный ряд обладает структурой, не совпадающей со структурой стандартного ряда теории возмущений. Достоинство перестроенного ряда заключается в том, что уже в низшем (четвертом) порядке теории возмущений он содержит в виде $\delta$-функции Дирака сингулярное слагаемое в стандартной теории отсутствующее. Это означает, что в условиях резонансного рассеяния сечение рассеяния обладает особенностью более высокого порядка, нежели следующее из стандартной теории. Сама возможность перестроения ряда указывает на то, что мы имеем дело с расходящимся рядом. Это обстоятельство не является новым. Расходимость рядов теории возмущений в квантовой теории хорошо известна. Ситуация усугубляется тем, что нам приходится работать в районе резонансной сингулярной точки. Нестандартность и неоднозначность построения решений в виде ряда демонстрируется равенством (23). Результат расчета формулы (23) зависит от порядка выполнения операций. Если, прежде всего, выполнить операцию суммирования по индексу $i$, то мы вернемся к результатам параграфа 3. Если же в формуле (23) предварительно выполнить операции квантового усреднения, а суммирование по индексу $i$ оставить напоследок, то получим формулу (24), содержащую в приближении $\propto q^4$ сингулярный член. Различие полученных таким образом решений и видимая абсурдность результата вынуждает выбрать одно из решений (11) или (21). Мы отдаем предпочтение решению (21), поскольку этот результат получен как в параграфе 6 по перестроенной теории возмущений, так и в параграфе 5 независимым способом, прямого отношения к теории возмущений не имеющим.

Таким образом, рассеяние электрона на нейтроне обладает «гигантским» резонансом, который стандартной теорией возмущений описан быть не может.

**СПИСОК ЛИТЕРАТУРЫ**

1. Wilson F.L. Fermi's Theory of Beta Decay. *American Journal of Physics* 36 (1968) 1150…1160.

2.Ландау Л.Д., Лифшиц Е.М. *Квантовая механика*. Наука, М.: 1974.

3. Векленко Б.А. Когерентные эффекты в некогерентном канале рассеяния резонансного излучения возбужденными атомами. *ЖЭТФ* 1939 (2011) 856…867.

4.Клаудер Дж., Сударшан Э. *Основы квантовой оптики*. МИР, М.: 1970.

5.Bertoni D., Reddy S., Rrapaj E. Electron –neutron scattering and transport properties of neutron stars. *arXiv*:1409.7750vI [nucl-th] 27 Sep 2014.

6. Dirac PAM. Quantum theory of the emission and absorption of radiation. *Proc. Roy. Soc*. 1927;A114:243-265.

7.Ахиезер А.И., Берестецкий В.Б. *Квантовая электродинамика*. Наука, М.: 1969.

8.Wick G.The Evaluation of the CollisionMatrix.*Phys.Rev*. 80 (1950)268...272.

**REFERENCES**

1. Wilson F.L. Fermi's Theory of Beta Decay. *American Journal of Physics* 36 (1968) 1150…1160.

2. Landau L.D, Lifshitz E.M, *Quantum Mechanics*, Moscow, Nauka, 1989.

3.Veklenko B.A. Coherent Effects in the Incoherent Channel of Resonant Radiation Scattering from Excited Atoms.*JETP* 112 (2011) 744…755.

4. Klauder, Sudarshan E.C.G. *Fundamentals of quantum optics*. Syracuse University W.A. Benjamin, INC. New York, Amsterdam1968.

5.Bertoni D., Reddy S., Rrapaj E. Electron –neutron scattering and transport properties of neutron stars. *arXiv*:1409.7750vI [nucl-th] 27 Sep 2014.

6. .Dirac PAM. Quantum theory of the emission and absorption of radiation.*Proc. Roy. Soc.* 1927;A114:243…265.

7.Akhiezer A.I., Berestetskii V.B. *KvantovajaElektrodinamika*. [Quantum Electrodynamics].Moskow.Nauka. 1969.7.

8. Wick G. The Evaluation of the Collision Matrix.*Phys.Rev*. 80 (1950)268…272.

**On the theory of electron scattering by neutrons**

*Vasil'eva A.A., Veklenko B.A.*

The scattering electro –neutron proses creates as electro-conductivity and thermo-conductivity of neutron starts. Present day's theories lead to the too low values of these parameters. In present paper we pay attention to the fact that at resonant or quasiresonant scattering the conventional perturbation theory is not sufficient. By the approaching to the resonance the increasing of the cross-section is too larger than the one predicted by conventional perturbation theory. In the resonance vicinity the pole peculiarity exists that can't be described by perturbation theory. The proposed mathematical formalism is taken from the theory of light –excited atom scattering. This formalism principally uses of the properties of quantum coherent states of electromagnetic fields. In present paper it is shown how one can use such formalism for describing the fermions scattering.

Key words: neutrons, electrons, Fermi Hamiltonian, perturbation theory, gigantic resonance.

*Васильева Анна Анатольевна*, аспир. ИОФАН ,*117312 Москва, Российская Федерация, ул. Вавилова дом 38.* Дом.адрес: 428028 г. Чебоксары, Проспект Тракторостроителей дом 71, кв. 80.

Тел. 8(8352)531279 (дом),  8(926)6388590 (моб). E-mail: anna_vasiljeva@bk.ru

*Векленко Борис Александрович*,д.ф-м наук, профессор, гл. научный сотрудник Объединенного Института Высоких Температур  (ОИВТ).  *125412, Москва, Российская Федерация, Ижорская ул., 13, стр. 2.* Дом.адрес: 107207 г. Москва, ул. Байкальская, дом 30, корп.2, кв.49.
Тел. 8(495)362-53-10 (сл.),  8(495)466-28-97 (дом), 8-962-932-87-52 (моб). E-mail: VeklenkoBA@yandex.ru